# Efeito de um Termo Dissipativo no Sistema Hamiltoniano de Ondas de Deriva

# (Effect of a Dissipative Term in the Drift Waves Hamiltonian System)

Ricardo S. Oyarzabal[1]; José D. Szezech Júnior[2]; Antonio M. Batista[2*]; Iberê L. Caldas[3]; Ricardo L. Viana[4], Kelly C. Iarosz[5]

1 Pós-Graduação em Ciências/Física, Universidade Estadual de Ponta Grossa, Paraná, Brasil
2 Departamento de Matemática e Estatística, Universidade Estadual de Ponta Grossa, Paraná, Brasil
3 Departamento de Física Aplicada, Instituto de Física, Universidade de São Paulo, São Paulo, Brasil
4 Departamento de Física, Universidade Federal do Paraná, Paraná, Brasil
5 Instituto de Física (Pós-Doutorado CNPq), Universidade de São Paulo, São Paulo, Brasil
*e-mail: antoniomarcosbatista@gmail.com

## Resumo

O presente trabalho analisa o modelo Hamiltoniano de ondas de deriva que descreve o transporte caótico de partículas no confinamento do plasma. Com apenas uma onda de deriva o sistema é integrável e apresenta órbitas regulares. Ao adicionarmos uma segunda onda, o sistema pode ou não ser integrável dependendo da velocidade de fase de cada onda. Com a mesma velocidade de fase o sistema ainda é integrável. Quando as velocidades de fase entre as duas ondas são diferentes, o sistema apresenta comportamento caótico. A este modelo acrescentamos uma pequena dissipação. Na presença de uma fraca dissipação, para certas condições iniciais, observamos órbitas transientes que convergem para atratores periódicos.

**Palavras-chave:** Plasma, ondas de deriva, sistemas Hamiltonianos, sistemas fracamente dissipativos.

## Abstract

This paper analyses the Hamiltonian model of drift waves which describes the chaotic transport of particles in the plasma confinement. With one drift wave the system is integrable and it presents stable orbits. When one wave is added the system may or may not be integrable depending on the phase of each wave velocity. If the two waves have the same phase velocity, the system is integrable. When the phase velocities between the two waves are different, the system shows chaotic behaviour. In this model we add a small dissipation. In the presence of a weak dissipation, for different initial conditions, we observe transient orbits which converge to periodic attractors.

**Key words:** plasma, drift waves, Hamiltonian systems, weakly dissipative systems.

## 1 Introdução

Plasma é definido como um gás quase-neutro formado por partículas carregadas e neutras que apresentam comportamento coletivo. No estado plasma os átomos estão dissociados em íons positivos e elétrons [1]. Apesar de ser raro no planeta Terra, o plasma ocorre naturalmente e abrange a maior parte do universo. Interiores e atmosferas estelares, nebulosas gasosas e hidrogênio interestelar são exemplos de matéria que se encontra neste estado [2]. Também encontra-se plasma em ventos solares, relâmpagos, gás em um tubo fluorescente e no escape de um foguete [3]. A ocorrência natural de plasmas em altas temperaturas é a razão de ser chamado de "o quarto estado da matéria" [4].

Existem três características que definem o plasma. A primeira, relativa ao comportamento coletivo, diz que partículas carregadas devem estar suficientemente próximas para que assim possam influenciar as demais partículas carregadas da vizinhança. A distância da interação das partículas carregadas é estimada pela esfera de Debye, cujo raio é definido como comprimento de Debye. A segunda característica do plasma é que o comprimento de Debye deve ser pequeno quando comparado ao tamanho físico do plasma. Desta forma o plasma é praticamente neutro. A terceira característica importante é que a frequência dos elétrons do plasma é alta quando comparada com as frequências de colisões entre os elétrons e as partículas neutras [1].

As propriedades de um plasma são estabelecidas pela interação entre as partículas, ou seja, a característica básica que define o comportamento de um plasma, o que o diferencia de um fluido ou um sólido, é seu efeito coletivo.

Devido a grande abrangência de forças eletromagnéticas, cada partícula carregada interage simultaneamente com um número elevado de outras partículas carregadas, resultando no comportamento coletivo, responsável pela riqueza de fenômenos físicos que ocorrem em um plasma [5]. Em um plasma as partículas carregadas geram concentração local de cargas positivas e negativas, então surgem campos elétricos. O movimento das cargas geram correntes elétricas, e consequentemente são produzidos campos magnéticos. Esses campos afetam o movimento a longo alcance de outras partículas carregadas [1].

Em 1952 surgiu a proposta de controle da reação de fusão da bomba de hidrogênio para criação de um reator [6]. A fusão termonuclear controlada é uma das aplicações da física de plasma. Para se construir reatores que gerem energia a partir das reações de fusão termonuclear controlada é necessário confinar o plasma a elevadas temperaturas e densidade em um campo magnético por um tempo suficientemente longo para que a energia produzida seja maior que a gasta com o confinamento e aquecimento [7].

Uma série de problemas prejudicam o confinamento do plasma, o que motiva várias pesquisas sobre o assunto, tais como turbulência eletrostática na borda do plasma [8, 9] e o transporte anômalo nas paredes do tokamak [10, 11]. O transporte anômalo é ocasionado pelo aparecimento de ondas de deriva e corresponde a uma manifestação da turbulência eletrostática na borda do plasma. As partículas escapam do confinamento dirigindo-se para as paredes do tokamak, desta forma ocorre o resfriamento do plasma e a perda de energia [11, 12].

Ondas de deriva são flutuações eletrostáticas que surgem devido a não-uniformidade do plasma. Estas ondas desempenham um papel importante no transporte de energia e de partículas confinadas magneticamente em um plasma. A presença de gradientes de densidade na borda do plasma originam turbulência, que podem explicar as taxas de transporte anômalo observados experimentalmente [13]. Um modelo de ondas de deriva foi proposto por W. Horton, nos anos 50 [13], ao estudar o transporte radial de partículas em um plasma confinado. Instabilidades surgem a partir de gradientes de densidade devido a estas ondas, então ocorre o transporte anômalo de partículas através do campo magnético de confinamento [14]. A manifestação de ondas de deriva, em plasmas magnetizados, é responsável pelo mecanismo dominante no transporte de partículas, de energia e na dinâmica que ocorre em todas as linhas do campo magnético [12]. Através do estudo das ondas de deriva é possível prever instabilidades que prejudicam o confinamento, bem como gradientes de densidade, gradientes de temperatura e efeitos do aprisionamento de partículas [15, 16].

Máquinas como tokamak são utilizadas para realizar experiências com plasmas confinados magneticamente. O tokamak, (toroidal'naya kamera magnitnymi katush-kami) que significa câmara de confinamento magnético toroidal, foi montada pela primeira vez nos anos 50 na Rússia [7]. O plasma é produzido em uma câmara toroidal e confinado pela superposição dos campos magnéticos toroidal e poloidal. A combinação destes campos possibilita o confinamento adequado do plasma em tokamaks. O campo magnético toroidal é obtido pela corrente elétrica que passa por espiras colocadas ao redor da câmara. A corrente elétrica do plasma produz o campo magnético poloidal.

A figura 1 mostra o esquema de um tokamak, aparelho que foi desenvolvido para estudar fenômenos de plasma. A corrente no primário de um transformador ôhmico, a qual o plasma representa o secundário produz a variação do fluxo magnético que gera campo elétrico na direção toroidal que cria a corrente de plasma. Existem espiras paralelas, que geram campos magnéticos verticais utilizados para controlar a posição e o formato da coluna de plasma. Essas, estão localizadas na parte superior e inferior da câmara toroidal.

A estrutura deste artigo é a seguinte: na seção II é feita uma breve introdução sobre sistema Hamiltoniano. Na seção III abordamos o modelo de ondas de deriva chegando na Hamiltoniana adimensional. O sistema integrável com apenas uma onda de deriva é relatado na seção IV. Acrescentamos uma segunda onda ao sistema na seção V e observamos as dinâmicas regular e caótica. Na seção VI adicionamos uma pequena quantidade de dissipação, onde podemos observar o surgimento de atratores. A seção VII é dedicada às conclusões e aos agradecimentos.

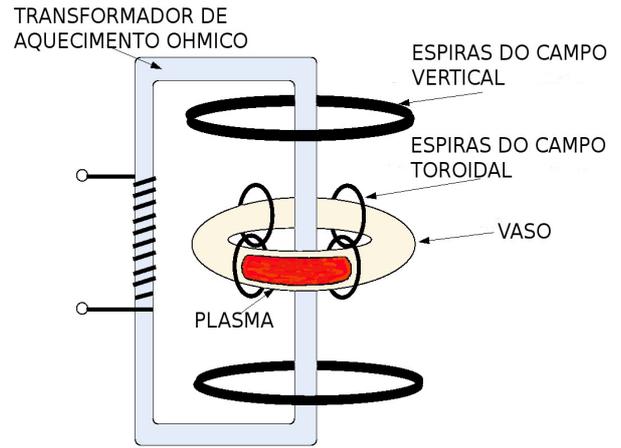

**Figura 1.** Esquema dos principais componentes do tokamak.

## 2 Sistema Hamiltoniano

Uma função Hamiltoniana combina equações diferenciais e princípios variacionais que podem ser escritos na forma de um sistema de equações. A função $H$ é chamada de Hamiltoniana, $t$ é o tempo e as variáveis $x$ e $y$ são as coordenadas canônicas. O estado do sistema é dado como o conjunto de pontos em $x$ e $y$ que formam o espaço de fase [25, 26].

Sistemas Hamiltonianos são sistemas dinâmicos conservativos que mantêm a energia constante a medida que o tempo evolui. O modelo de ondas de deriva é um sistema Hamiltoniano, que no plano é definido como um sistema de equações diferenciais com um grau de liberdade,

$$\frac{dx}{dt}=\frac{-\partial H}{\partial y},$$
$$\frac{dy}{dt}=\frac{\partial H}{\partial x}, \qquad (1)$$

onde $H(x, y)$ é uma função contínua e diferenciável.

Um sistema Hamiltoniano pode ser definido como

$$H(x,y)=K(x,y)+V(x,y), \qquad (2)$$

onde K é a energia cinética e V é a energia potencial.

A energia total $H(x, y)$ é uma constante do movimento de acordo como o princípio da conservação de energia [27]. A derivada total ao longo da trajetória é, utilizando a equação (1),

$$\frac{dH}{dt}=\frac{\partial H}{\partial x}\frac{dx}{dt}+\frac{\partial H}{\partial y}\frac{dy}{dt}=0. \qquad (3)$$

Portanto, $H(x, y)$ é constante ao longo de curvas que são soluções de (1) e suas trajetórias ficam em contornos definidos por $H(x, y) = C$, onde $C$ é uma constante. Sistemas Hamiltonianos com 1 grau de liberdade são sempre integráveis.

Um sistema em que não existem forças externas ou dissipativas é possível distinguir dois tipos de dinâmica, regular ou caótica [28]. O comportamento regular ocorre em ilhas KAM, toros KAM, submersos em um mar caótico. Isso acontece para sistemas de dois e 1 grau e meio de liberdade. O "Teorema de Kolmogorov Arnold e

Moser" (KAM) é um resultado em sistemas dinâmicos sobre a persistência de movimentos quase-periódicos sob pequenas perturbações [18, 28]. Em um sistema hamiltoniano integrável a trajetória no espaço de fase é uma superfície chamada toro invariante. O movimento em um torus pode ser generalizado como um toro no espaço de fase [17, 18, 21].

Em sistemas integráveis perturbados, diferentes condições iniciais irão traçar diferentes superfícies invariantes. As órbitas deste sistema são quase periódicas e tais movimentos persistem mesmo sob uma pequena perturbação. Este movimento quase periódico possui um número finito de frequências incomensuráveis. A análise de estabilidade ´e o tema da teoria KAM. Em cada uma destas curvas, todos os pontos se movem com a mesma velocidade de rotação, o chamado número de rotação [18].

## 3  Modelo de ondas de deriva

No modelo ondas de deriva, consideramos o movimento em campos elétrico e magnético de partículas carregadas em um tokamak devido à presença de uma ou mais ondas de deriva [13]. Em um plasma confinado, as trajetórias destas partículas estão sujeitas ao campo eletromagnético e podem ser determinadas pela força de Lorentz

$$m\frac{d\vec{v}}{dt}=q(\vec{E}+\vec{v}\times\vec{B}), \quad (4)$$

onde $m$ é a massa, $q$ é a carga, $v$ é a velocidade, $\vec{E}$ é o campo elétrico e $\vec{B}$ o campo magnético. Neste modelo descreve-se o movimento de deriva por meio de um centro de guia. O ponto cujo lugar geométrico é o centro do movimento circular é chamado de centro de guia. Assim, a velocidade no centro de guia é dada por

$$\vec{v}_E=\frac{\vec{E}\times\vec{B}}{B^2}, \quad (5)$$

onde $\vec{v}_E$ é a velocidade de deriva do centro de guia devido aos campos $\vec{E}$ e $\vec{B}$.

Um plasma confinado possui coordenadas toroidais, mas é possível simplificar a geometria do sistema com uma aproximação para coordenadas cartesianas. Primeiramente reduz-se o toroide para um cilindro e este para um plano. Isto é possível porque o raio do toróide é muito menor que o raio do plasma. A figura 2 ilustra o processo onde o cilindro é aproximado para um plano. Considera-se as distâncias $ad$ e $bc$ muito menores que o comprimento $ab$ e $cd$. Desta forma, tal aproximação só é válida para as bordas do plasma.

Definimos a direção radial $r$ de um cilindro como o eixo $x$ do sistema cartesiano, a direção poloidal $\theta$ como eixo $y$ e a direção toroidal $z$ do cilindro como eixo $z$ [8].

O campo elétrico é escrito a partir do gradiente do potencial $\Phi$

$$\vec{E}=-\nabla\Phi, \quad (6)$$

e o campo magnético uniforme como

$$\vec{B}=-B_0\hat{e}_z, \quad (7)$$

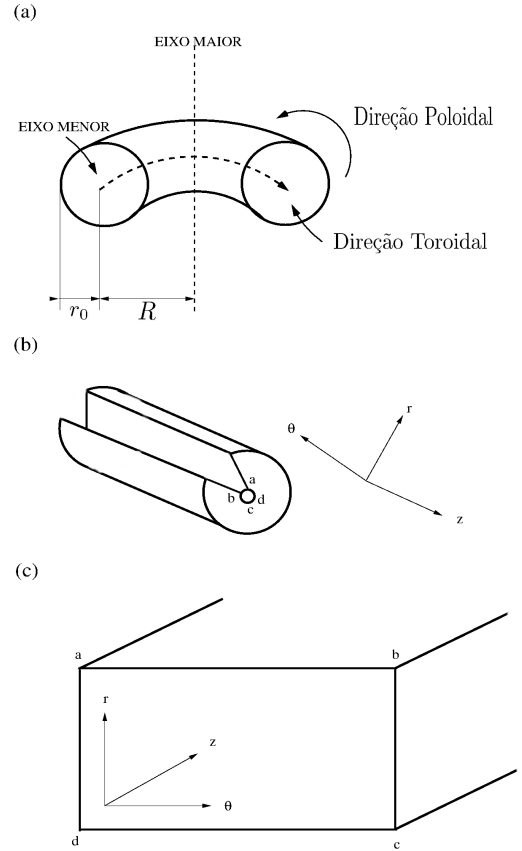

**Figura 2** - Esquema para transformação de coordenada toroidal para cartesiana.

onde $\hat{e}_z$ é o vetor unitário na direção de $z$ e $B_0$ a intensidade do campo magnético toroidal de equilíbrio.

A velocidade de deriva pode ser escrita como

$$\vec{v}_E=\frac{1}{B_0}\nabla\Phi\times\hat{e}_z, \quad (8)$$

$$\vec{v}_E=\frac{-1}{B_0}\frac{\partial\Phi}{\partial y}\hat{e}_z \quad (9)$$

A Hamiltoniana é dada por

$$H=\frac{\Phi}{B_0}, \quad (10)$$

onde consideramos $x$ o momento generalizado e $y$ a coordenada generalizada.

O sistema de equações (1) possui um par canônico de coordenadas (x, y) e um grau de liberdade, portanto será integrável [8]. Consideramos agora um modelo para N ondas de deriva [13] onde as ondas se propagam na direção poloidal $y$ com uma modulação na direção radial $x$. Neste modelo tem-se uma plasma não uniforme, magnetizado e sujeito a um campo elétrico radial constante $\vec{E}=E_0\hat{e}_x$. O potencial elétrico pode ser escrito na forma

$$\Phi(x,y,t) = \Phi_0(x) + \sum_{i=1}^{N} A_i \, sen(k_{xi}x) \cos(k_{yi}y - w_i t), \quad (11)$$

onde $A$ é a amplitude de cada onda, $k_{xi}$ e $k_{yi}$ são números de onda na direção $x$ e $y$, $\omega_i$ é a frequência angular da i-ésima onda e $\Phi_0$ é o potencial do plasma em equilíbrio. O potencial $\Phi(x,y,t)$ é a superposição dos potenciais de equilíbrio $\Phi_0(x)$ e de $\Phi_1(x, y-\omega t)$. Este potencial é a superposição de ondas eletrostáticas, ondas de deriva, com propagação na direção poloidal $y$ e modulação na direção radial $x$ [7].

A Hamiltoniana para $N$ ondas estacionárias é

$$H(x,y,t) = \frac{1}{B_0}[\Phi_0(x) + \sum_{i=1}^{N} A_i \, sen(k_{xi}x) \cos(k_{yi}y - w_i t)] \quad (12)$$

Adotamos o potencial elétrico na forma linear

$$\Phi_0(x) = ax, \quad (13)$$

onde $a$ é o campo elétrico constante na direção de $x$ [8]. A Hamiltoniana escrita na forma adimensional é

$$H(x,y,t) = ax + \sum_{i=1}^{N} A_i \, sen(k_{xi}x) \cos(k_{yi}y - w_i t) \quad (14)$$

A Hamiltoniana (14) é integrável para N = 1 com existência de solução analítica e trajetórias regulares.

Para N > 1, o sistema é integrável se todas as $N$ ondas tiverem a mesma velocidade de fase. Se pelo menos uma onda tiver velocidade de fase diferente das demais, o sistema não será integrável e algumas trajetórias serão caóticas [13].

## 4 Sistema integrável com uma onda de deriva

Para o modelo de ondas de deriva, primeiramente consideramos o sistema integrável com apenas uma onda de deriva. Este sistema Hamiltoniano depende do tempo, mas é possível fazer uma mudança de referencial da onda que elimina o tempo da equação (transformação canônica).

A Hamiltoniana para N = 1 em termo das novas variáveis é

$$H(x,y,t) = (a-u_1)x + A_i \, sen(k_{xi}x) \cos(k_{yi}y), \quad (15)$$

onde $u_1 = \frac{w_1}{k_{y1}}$ é a velocidade de fase da onda.

Aplicando as equações de Hamilton (1) é possível chegar ao sistema normalizado

$$\begin{aligned} \frac{dx}{dt} &= sen(x) sen(y), \\ \frac{dy}{dt} &= U + \cos(x)\cos(y), \end{aligned} \quad (16)$$

onde $x = k_x x$ e $y = k_y y$ e $U$ é o parâmetro de confinamento definido como

$$U = \frac{a - u_1}{A_1 k_{xi}}, \quad (17)$$

que influencia nas trajetórias das partículas e caracteriza no espaço de fase a forma das linhas de fluxo descritas pela Hamiltoniana. Este parâmetro possui dependência com o campo elétrico da descarga no tokamak e dos demais parâmetros da onda de deriva [7, 9].

Agora, integramos numericamente o sistema (16). A figura 3 mostra a trajetória das partículas para o sistema de equações (16) com os eixos $x$ e $y$ normalizados e com o parâmetro de confinamento $U = 0$ e $U = 0,5$.

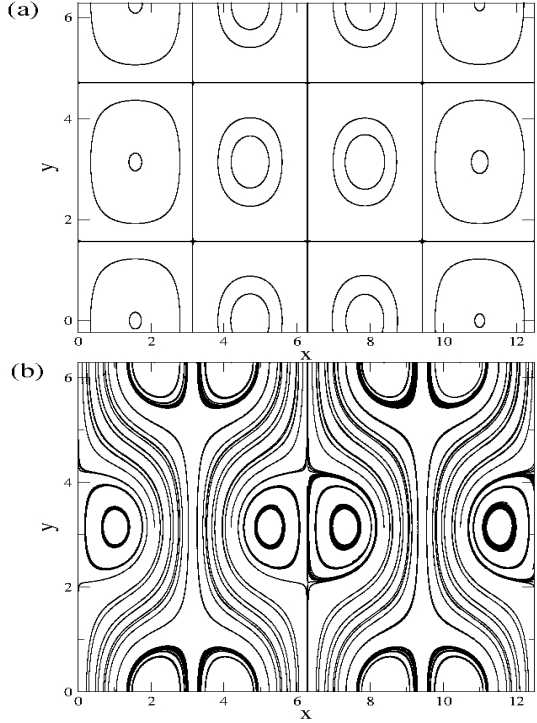

**Figura 3** - Espaço de fase para uma onda de deriva com o parâmetro de confinamento (a) U = 0 e (b) U = 0, 5.

Para $U = 0$ as trajetórias são fechadas e as partículas confinadas nessas trajetórias giram em torno de pontos elípticos. As trajetórias confinadas estão em ilhas, curvas fechadas. As linhas são as separatrizes, cujo retângulo é chamado de célula. Nesse caso não existe nenhuma trajetória entre células, as partículas ficam confinadas e não existe transporte de partículas em nenhuma direção. As separatrizes se interceptam em um ponto hiperbólico de sela. Em um ponto hiperbólico de sela as trajetórias aproximam-se do ponto de equilíbrio por uma das direções e afastam-se por outra, desta forma um ponto de sela é sempre instável [29].

Para o parâmetro de confinamento $U = 0,5$ é possível notar que além das ilhas surgem trajetórias abertas, não confinadas, que funcionam com barreiras de transporte na direção de $x$. Isto impede as partículas de migrarem ao longo da direção radial.

## 5 Sistema perturbado com duas ondas de deriva

O sistema com duas ondas de deriva ainda pode ser integrável se a velocidade de fase das ondas for igual a velocidade de deriva do campo elétrico ($U = 0$). Adota-se o referencial da primeira onda, então a Hamiltoniana para duas ondas de deriva (N = 2) torna-se

$$H(x,y,t) = (a-u_1)x + A_1 sen(K_{x1}x)\cos(K_{y1}y) \\ + A_2 sen(K_{x2}x)\cos(k_{y2}(y-ut)), \quad (18)$$

onde $u_1 = \dfrac{w_2}{k_{y2}} - \dfrac{w_1}{k_{y1}}$ é a diferença de fase entre as duas ondas.

Com a variável tempo o sistema ficará com um grau e meio de liberdade. Através de (18) obtemos o sistema

$$\dfrac{dx}{dt} = A_1 k_{y1} sen(K_{x1}x) sen(k_{y1}y) + A_2 k_{y2} sen(k_{x2}x) \\ sen(k_{y2}(y-ut))$$

$$\dfrac{dy}{dt} = (a-u_1) + A_1 k_{x1} \cos(K_{x1}x)\cos(k_{y1}y) + A_2 k_{x2} \\ \cos(k_{x2}x)\cos(k_{y2}(y-ut)). \quad (19)$$

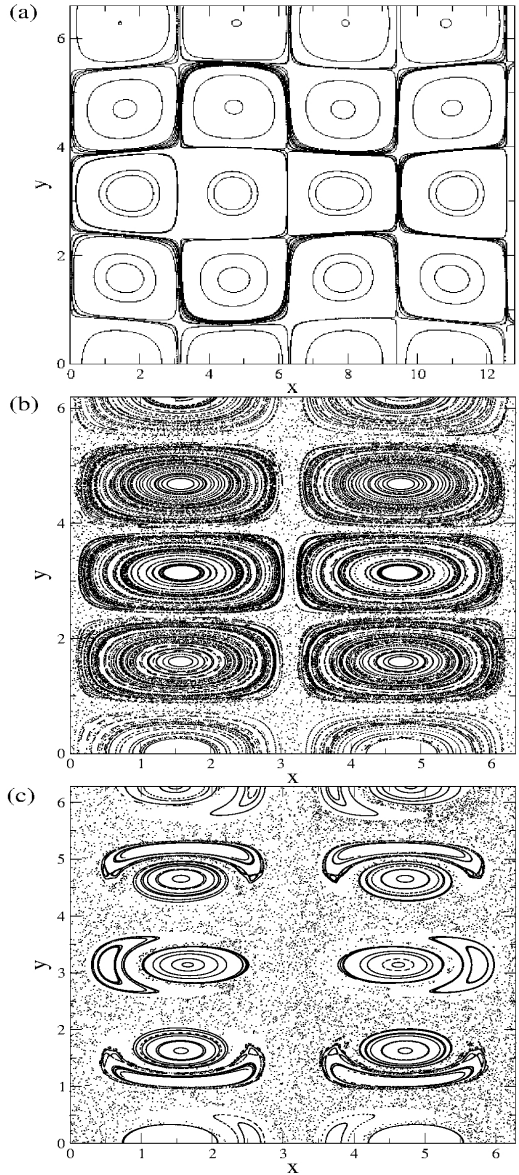

**Figura 4** – Simulação numérica para os parâmetros $a$-$u_1$ = 0, $A_1$ = 1, $A_2$ = 0, 1, $k_{y1}$ = 1, $k_{y2}$ = 0, 5, $k_{x1}$ = 1, $k_{x2}$ = 3, (a) $u = 0$ (b) $u = 0, 5$, (c) $u = 1, 5$.

As trajetórias são exibidas no plano $xy$ em mapas de Poincaré, que é a interseção de uma órbita periódica e caóticas no espaço de um sistema dinâmico contínuo com uma certo subespaço de dimensão menor, chamado a seção de Poincaré [29]. Como a Hamiltoniana (18) é periódica, fazemos a seção de Poincaré em tempos múltiplos de ($2\pi/u$), assim o conjunto de pontos com tempo múltiplo deste período formam um mapa.

Os parâmetros utilizados são: $A_1 = 1$, $A_2 = 0, 1$, $k_{y1} = 1$, $k_{y2} = 0, 5$, $k_{x1} = 1$, $k_{x2} = 3$ [7]. Neste trabalho utilizamos a razão do parâmetro $k_{x2}/k_{x1}$ irracional, assim não existe barreira de transporte na direção de $x$ [8]. Existem trajetórias que não são confinadas, e as partículas podem ir de uma ilha para outra, mesmo para o caso de $u = 0$ [7, 8]. Para estudar somente a interação entre as duas ondas considera-se o parâmetro de confinamento $U = 0$, para isto, escolhe-se a velocidade de fase igual a velocidade de deriva do campo elétrico $a-u_1 = 0$.

A figura 4 mostra o resultado de algumas simulações numéricas para diferentes valores do parâmetro $u$. Em (a) temos o espaço de fase para o parâmetro $u = 0$, as duas ondas com mesma velocidade de fase. É possível observar que as trajetórias são regulares, o sistema é integrável. Em (b) temos o parâmetro $u = 0,5$ e a Hamiltoniana quasi-integrável, ou seja, coexiste regiões regulares e caóticas no espaço de fase. Como a segunda onda tem amplitude pequena $A_2 = 0,1$, apenas as trajetórias próximas à separatriz são caóticas. Em (c) aumentamos a diferença da velocidade de fase para $u = 1,5$, onde observa-se órbitas quasi-periódicas e caos. Caos determinístico é a existência de comportamento irregular, ou aperiódico, com dependência sensível das condições iniciais, ou seja, dois processos originados por condições iniciais ligeiramente diferentes divergem exponencialmente com o tempo [30].

## 6 Sistemas dissipativos

A presença de perturbações externas, como dissipação aparece em vários contextos físicos e existem muitos trabalhos sobre este tema [19, 20]. Quando acrescentamos uma dissipação com uma magnitude muito pequena, o sistema passa a exibir um comportamento que é dominado principalmente pelo aparecimento de atratores periódicos de diferentes períodos [21]. Um pequeno termo dissipativo adicionado ao sistema faz com que as órbitas periódicas do mapa de Poincaré tornem-se um atrator e os toros quasi-estáveis desapareçam para dar lugar às bacias dos atratores [22]. Bacias de atração de um sistema fornecem o conjunto de condições iniciais que evoluem para determinados atratores após um transiente [23, 24].

Com o acréscimo de uma pequena quantidade de dissipação no sistema, órbitas periódicas tornam-se bacias de atração e o movimento caótico é substituído por um extenso transiente caótico que ocorre antes que a trajetória se estabeleça em uma desta bacias [17]. O estado final de um sistema depende deste processo intermediário, onde o objeto de análise não é somente o comportamento final, mas sim a complexa dinâmica transiente que conduz a este resultado [31].

Um subconjunto $X$ do espaço de fase é um atrator se: $X$ é invariante sob um fluxo, existe uma vizinhança aberta em torno de $X$ que converge a $X$ sob tal fluxo, nenhuma parte de $X$ é transiente e $X$ não pode ser decomposto em duas partes não sobrepostas invariantes. Assim, a bacia de atração de $X$ é um conjunto de estados no espaço de fase que se aproxima de $X$ para um tempo que tende ao infinito [17].

Do sistema de equações (19), modelo de ondas de deriva, acrescentamos um termo dissipativo $v$ em uma das equações. Desta forma obtemos o sistema

$$\frac{dx}{dt} = A_1 k_{y1} sen(K_{x1} x) sen(k_{y1} y) + A_2 k_{y2} sen(k_{x2} x) sen(k_{y2}(y-ut))$$

$$\frac{dy}{dt} = (a-u_1) + A_1 k_{x1} cos(K_{x1} x) cos(k_{y1} y) + A_2 k_{x2} cos(k_{x2} x) cos(k_{y2}(y-ut)) - vy, \quad (20)$$

onde $v$ é a dissipação.

Os parâmetros que utilizamos na simulação numérica são os mesmos utilizados na seção anterior: $a - u_1 = 0$, $A_1 = 1$, $A_2 = 0, 1$, $k_{y1} = 1$, $k_{y2} = 0, 5$, $k_{x1} = 1$, $k_{x2} = 3$. A diferença de fase das duas ondas $u = 1, 5$ é a mesma da figura 4(c), agora com a introdução de um termo dissipativo no valor de $1 \times 10^{-4}$.

Para visualizarmos melhor algumas órbitas quase-periódicas do sistema, inicialmente fazemos a simulação numérica para o sistema conservativo, figura 5. Consideramos $v = 0$, em uma grade no intervalo [0, 4; 3, 0] no eixo x e [0, 9; 2, 1] no eixo y. Na ampliação é possível observar órbitas quase-periódicas, ilhas KAM, que são representadas pelas linhas pretas e estão dispersas em um mar caótico. A figura 5 mostra a simulação numérica com a dissipação. O conjunto de condições inicias que estão dentro desta grade podem convergir para três atratores diferentes, representados pelo símbolo X nas cores azul, verde e vermelho. Para este valor de dissipação, o conjunto de órbitas em torno $x \approx 1, 60$ e $y \approx 1, 65$ convergem para o atrator representado por X em azul. Este é um atrator de período um. Outro atrator, também de período 1, é representado na figura 5 por X em verde. O conjunto de órbitas em torno $x \approx 1, 75$ e $y \approx 1, 05$ convergem, depois de um tempo muito longo, para este atrator, bem como pontos que se encontram em órbitas caóticas. Certas condições iniciais do sistema convergem para um atrator que é representado com o símbolo X em vermelho. Este é um atrator de período 10. Observa-se que este atrator é oriundo de pequenas ilhas que estão fora das órbitas maiores que convergem para o atrator representado em verde. Somente pontos que estão dentro destas pequenas ilhas convergem para este atrator.

## 7 Conclusões

Neste trabalho investigamos a interação entre duas ondas de deriva que surgem na borda do plasma em tokamaks. O modelo utilizado neste trabalho é Hamiltoniano, onde existe movimento regular em torno de órbitas estáveis, bem como círculos invariantes chamados de ilhas KAM. A dinâmica de tal sistema está localizada no torus $[0, 2\pi] \times [0, 2\pi]$, portanto, órbitas periódicas que são soluções das equações residem no torus.

Quando acrescentamos dissipação ao sistema, o espaço de fases já não se limita ao torus. O movimento acontece no cilindro $[0, 2\pi] \times R$. A dissipação conduz a uma separação da sobreposição de órbitas periódicas pertencentes a uma determinada família com o aumento do módulo da velocidade no cilindro. Com uma dissipação de $1 \times 10^{-4}$ no sistema as condições iniciais convergem para atratores após um tempo transiente.

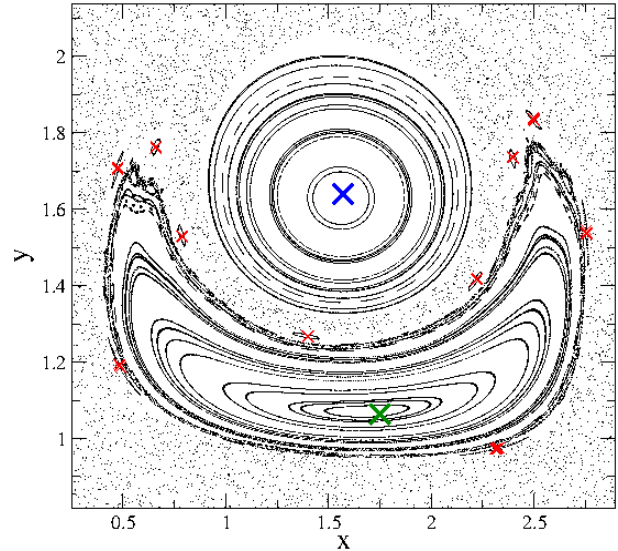

**Figura 5** - Simulação o numérica para duas ondas de deriva com os parâmetros $a - u_1 = 0$, $A_1 = 1$, $A_2 = 0, 1$, $k_{y1} = 1$, $k_{y2} = 0, 5$, $k_{x1} = 1$, $k_{x2} = 3$, $u = 1, 5$. Os atratores são representados por X nas cores azul, verde e vermelho.

## Agradecimentos

## Referências